\documentclass[a4paper,11pt]{article}
\usepackage{pos}

\title{Cross-calibration and combined analysis of the CTA-LST prototype and the MAGIC telescopes}
 \ShortTitle{MAGIC-LST cross-calibration and combined analysis}

\author*[a]{Y.~Ohtani}
\author[b,c]{A.~Berti}
\author[c]{D.~Depaoli}
\author[c]{F.~Di Pierro}
\author[b]{D.~Green}
\author[b]{L.~Heckmann}
\author[b]{M.~Hütten}
\author[a]{T.~Inada}
\author[d]{R.~López-Coto}
\author[c]{E.~Medina}
\author[e]{A.~Moralejo}
\author[f]{D.~Morcuende}
\author[c]{G.~Pirola}
\author[a]{M.~Strzys}
\author[g]{Y.~Suda}
\author[a]{I.~Vovk}

\affiliation[a]{ICRR, the University of Tokyo, 1-5-1 Kashiwa-no-ha, Kashiwa City, Chiba, Japan}
\affiliation[b]{Max Planck Institute for Physics, Föhringer Ring 6, 80805 München, Germany}
\affiliation[c]{INFN Sezione di Torino, Via Pietro Giuria, 1, 10125 Torino, Italy}
\affiliation[d]{INFN Sezione di Padova, Via Francesco Marzolo, 8, 35131 Padova, Italy}
\affiliation[e]{IFAE, Edifici Cn, Campus Universitat Autònoma de Barcelona, 08193 Barcelona, Spain}
\affiliation[f]{Universidad Complutense de Madrid, Av. Séneca, 2, 28040 Madrid, Spain}
\affiliation[g]{Physics Program, Graduate School of Advanced Science and Engineering, Hiroshima University, 739-8526 Hiroshima, Japan}

\forColl{CTA LST project and MAGIC} 

\emailAdd{ohtani@icrr.u-tokyo.ac.jp}

\abstract{The Cherenkov Telescope Array (CTA) will be the next generation gamma-ray observatory, which will consist of three kinds of telescopes of different sizes. Among those, the Large Size Telescope (LST) will be the most sensitive in the low energy range starting from 20~GeV. The prototype LST (LST-1) proposed for CTA was inaugurated in October 2018 in the northern hemisphere site, La Palma (Spain), and is currently in its commissioning phase. MAGIC is a system of two gamma-ray Cherenkov telescopes of the current generation, located approximately 100~m away from LST-1, that have been operating in stereoscopic mode since 2009. Since LST-1 and MAGIC can observe the same air shower events, we can compare the brightness of showers, estimated energies of gamma rays, and other parameters event by event, which can be used to cross-calibrate the telescopes. Ultimately, by performing combined analyses of the events triggering the three telescopes, we can reconstruct the shower geometry more accurately, leading to better energy and angular resolutions, and a better discrimination of the background showers initiated by cosmic rays. For that purpose, as part of the commissioning of LST-1, we performed joint observations of established gamma-ray sources with LST-1 and MAGIC. Also, we have developed Monte Carlo simulations for such joint observations and an analysis pipeline which finds event coincidence in the offline analysis based on their timestamps. In this work, we present the first detection of an astronomical source, the Crab Nebula, with combined observation of LST-1 and MAGIC. Moreover, we show results of the inter-telescope cross-calibration obtained using Crab Nebula data taken during joint observations with LST-1 and MAGIC.}

\FullConference{37$^{\rm{th}}$ International Cosmic Ray Conference (ICRC 2021)\\
		July 12th -- 23rd, 2021\\
		Online -- Berlin, Germany}

\begin{document}
\maketitle

\section{Introduction}
The Cherenkov Telescope Array (CTA) will be the next generation ground-based gamma-ray observatory, which will cover the energy range from 20~GeV to 300~TeV \cite{CTA}. There will be two observation sites in the northern and southern hemispheres in order to cover the entire sky. CTA will consist of three kinds of Cherenkov telescopes of different sizes, each of which is dominant in a different energy range. Among those, the Large Size Telescope (LST) is designed to be most sensitive in the lowest energy range starting from 20~GeV, with a large area of reflective mirrors of 23 m diameter. The prototype LST (LST-1) proposed for CTA was inaugurated in October 2018 in the northern site, La Palma (Spain), and is currently in its commissioning phase \cite{LSTcom}.\par
MAGIC is a system of current generation Cherenkov telescopes with two 17~m diameter mirrors, and can detect gamma rays between 50~GeV to 50~TeV \cite{MAGIChard} \cite{MAGICper}. The telescopes have been operating in stereoscopic mode since 2009 in La Palma and are located in the direct vicinity of LST-1. Since the distance between LST-1 and MAGIC is approximately 100~m, which is smaller than the typical $\sim$200~m diameter of the Cherenkov light pool, both systems can observe the same air shower events. Therefore, it is possible to cross-calibrate the telescopes by comparing the brightness of showers, estimated energies of gamma rays and the other parameters event by event. Ultimately, by performing the combined analyses of the events triggering the three telescopes, the shower geometry can be reconstructed more accurately, leading to better energy and angular resolutions, and a better discrimination of the background showers initiated by cosmic rays.\par
For that purpose, as part of the commissioning of LST-1, we performed joint observations of established gamma-ray sources with LST-1 and MAGIC. Also, we have developed a pipeline and Monte Carlo (MC) simulations dedicated to the analysis of the joint observation data. In this work, we present the analysis techniques and the results of the cross-calibration and combined analysis obtained using the Crab Nebula data taken during the joint observations with LST-1 and MAGIC.

\section{Event coincidence with timestamps}
Since LST-1 and MAGIC trigger and readout systems are independent, air shower events are recorded by both systems with different trigger rate and individual timestamps. We have developed a coincidence algorithm to find the events triggering both systems in the offline analysis based on the timestamps. In this algorithm, we select the following two time-related parameters. The first one is the "time offset", which compensates for the difference of the timestamps between the systems due to the geometry of the telescope array and/or any systematical reason. The other one is the full width of "coincidence window", which represents the time interval defined by the LST-1 events. The MAGIC-stereo events triggering the two telescopes and falling within the coincidence window are recognized as coincident events.\par
\begin{figure}
  \centering
  \includegraphics[width=15cm]{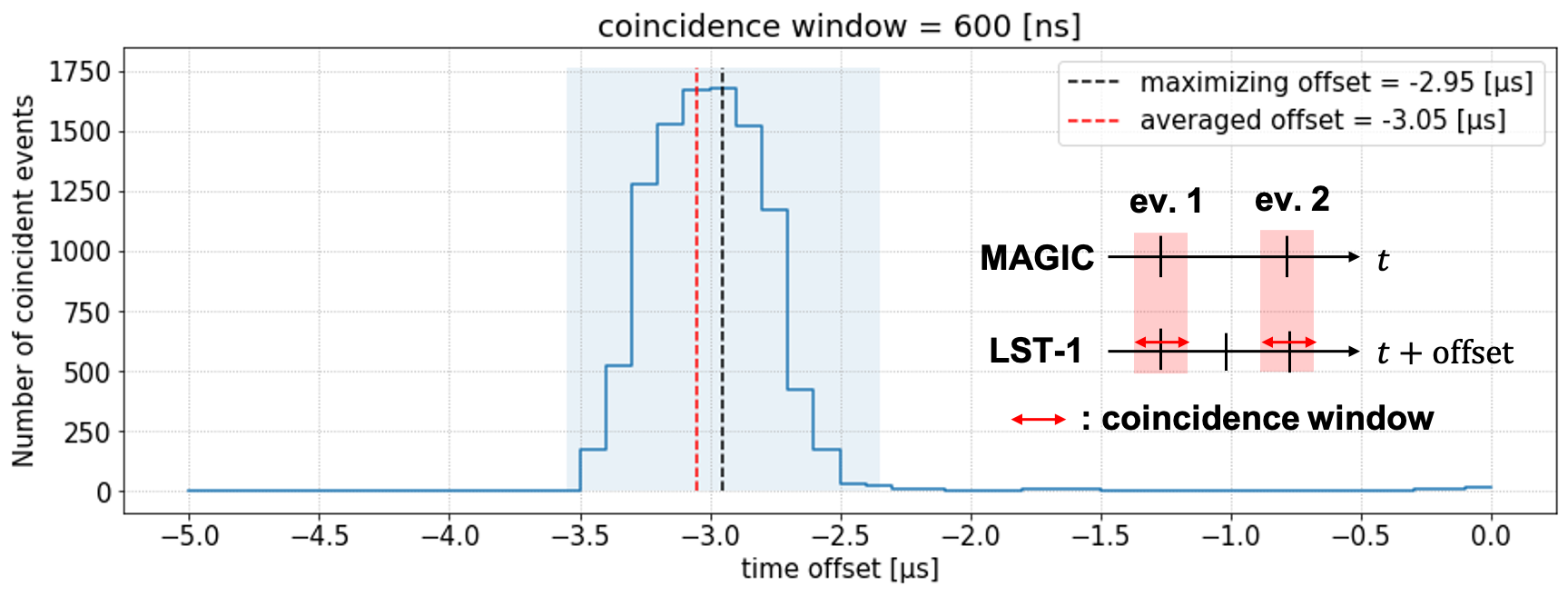}
  \caption{Example of searching for the coincident events with the timestamps of LST-1 and MAGIC-I. The width of the coincidence window is set to 600~ns. The black-dashed line represents the offset which maximizes the number of coincident events. The red-dashed line represents the averaged offset computed weighting with the number of coincident events within the region defined around the maximizing offset, which is shown as the blue-shaded area. See the text for more details.}
  \label{coincidence}
\end{figure}
In order to decide the best time offset parameter, we performed the following steps. At first, the number of coincident events is scanned for a given time offset value with a fixed width of the coincidence window. An example of the scan is shown in Fig.~\ref{coincidence}, where a small sample of the joint observation data is used. The distribution shows a clear peaked structure with a width comparable with the coincidence window. Next, two time offset values are investigated as candidates for the best parameter. One is the "maximizing offset" defined as the offset where the number of coincident events is maximized, and the other is the "averaged offset" computed weighting with the number of coincident events within a region around the peak, whose center is the maximizing offset and whose width is the coincidence window. Then, we investigated the dependence of these time offsets on the width of the coincidence window, and confirmed that the averaged offset shows stable behavior while the maximizing offset clearly changes its value. Thus, we decided to use the averaged offset as the best time offset since it well represents the feature of the distribution without depending on the width of the coincidence window.\par
Based on the averaged offset, we also optimized the width of the coincidence window in order to minimize the number of chance coincident events while retaining most of the number of actual coincident events. We checked the number of coincident events at the averaged time offset for each width of the coincidence window. We confirmed that there is a significant decrease in the number of coincident events for widths smaller than 600~ns, while above this value the number of chance coincidence events gradually increases. Thus, we decided to use the width of 600~ns as the optimized value of coincidence window. Considering the typical trigger rates of the telescope systems, i.e., 5~kHz for LST-1 and 300~Hz for MAGIC, the chance coincidence rate with 600~ns width of the coincidence window is estimated to be $\sim$1~Hz (= 5~kHz $\times$ 300~Hz $\times$ 600~ns). It should be noted that, for the validation of the algorithm, we confirmed that the coincidence rate of the cosmic-ray events obtained with the algorithm is consistent with the rate expected from MC simulation.\par
We investigated the reason why the 600~ns is plausible value for the width of the coincidence window. In principle, the most of the events triggering the Cherenkov telescopes have hadronic origin and isotropic distribution in the FoV. The variety of the off axis angles between the arrival direction and the line of sight of the telescopes makes a jitter for the timestamps of each telescope system. We analytically calculated the degree of the jitter considering the geometry of the telescope array and the pointing direction of the test sample used in Fig.~\ref{coincidence}, and found that it is around $\pm$16~ns. Since it is much smaller than the optimized width 600~ns, it cannot be responsible for the width of the coincidence window. On the other hand, considering that the precision on the timestamps are $\pm$100 ns for LST-1 and $\pm$200 ns for MAGIC \cite{MAGIChard} \cite{TIB}, the precision on the timestamp difference is $\sqrt{100^2 + 200^2}$ = 224 ns, and the width of 600 ns is the interval that includes 99\% of the coincident events.Thus, we concluded that the width 600~ns comes from the precision of the timestamps.

\section{Inter-telescope cross-calibration}
In this section, we present the results of the inter-telescope cross-calibration by using Crab Nebula observation data. We use data taken on 17th January 2020 with exposure of roughly 1.5 hours at low zenith angles from 7 to 20 degrees. LST-1 observed the source with the so-called ON mode, where the source is located on the center of the FoV, while MAGIC observed with the so-called wobble mode with an offset of 0.4 degrees from the center of the FoV \cite{obsmode}. Despite the difference of the observation mode, both systems could observe the same air shower events and so the data can be used for the cross-calibration.\par
The data are at first analyzed independently in each telescope system with corresponding analysis pipelines. The LST-1 data are analyzed with \texttt{cta-lstchain}, which is developed for the LST data analysis based on \texttt{ctapipe} \cite{ctapipe}. Also, we use the \texttt{CORSIKA} and \texttt{sim\_telarray} packages for the MC simulation of the air shower development and the telescope response \cite{corsim}. We implemented realistic hardware parameters obtained from the LST-1 commissioning into the sim\_telarray package to reproduce the actual telescope performance in the simulation. Especially the telescope light efficiency used in the simulation is tuned to match the one of the real data obtained with the muon analysis \cite{LSTper}. The reconstruction of the gamma-ray energy, arrival direction and the gamma-hadron separation is performed with the Random Forest (RF) method, which is trained by feeding the MC data described above. It should be noted that, for the purpose of the cross-calibration, the LST-1 event reconstruction is performed using only the shower images obtained by LST-1, not including any MAGIC information. The MAGIC data are, on the other hand, analyzed with MAGIC Analysis and Reconstruction Software (MARS) and MAGIC standard MCs \cite{MARS} \cite{MMC}. It also uses the RF method for the event reconstruction and the gamma-hadron separation \cite{RFmars}. Since MAGIC consists of two telescopes, the event reconstruction is performed not only using the image parameters but also the reconstructed geometrical shower parameters.\par
\begin{figure}
  \centering
  \includegraphics[width=15cm]{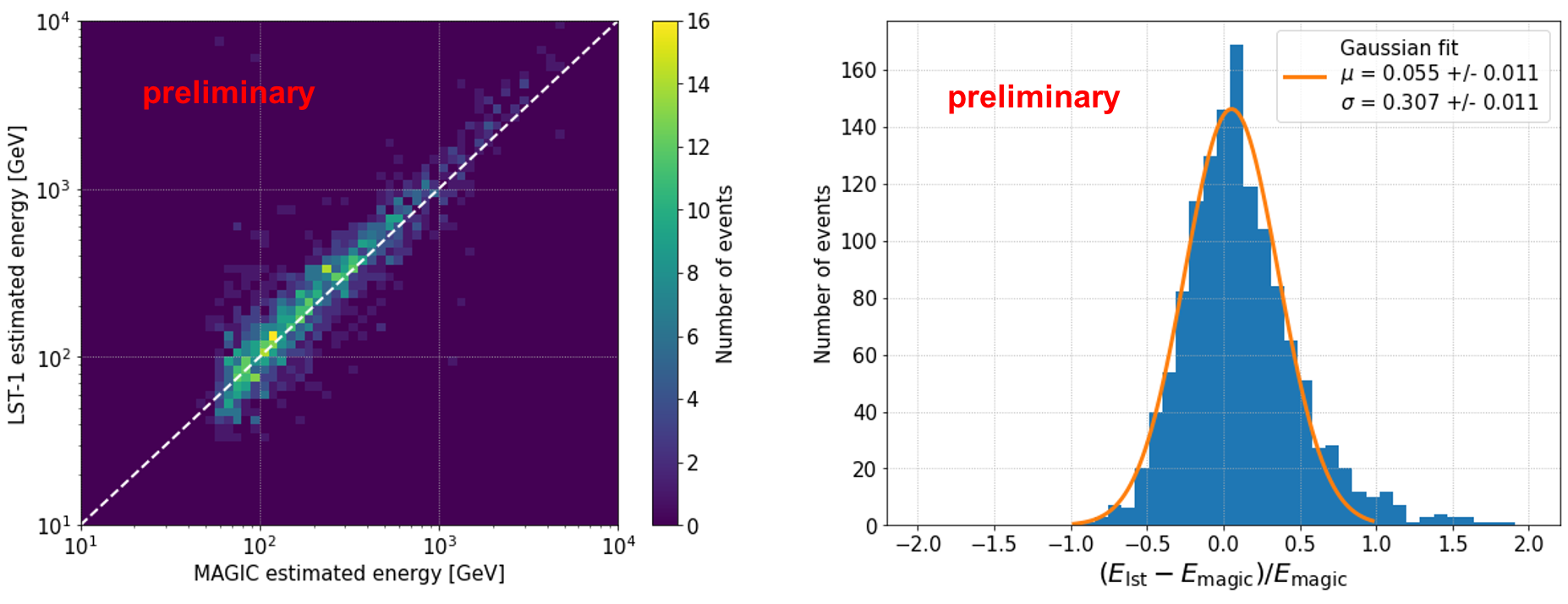}
  \caption{Comparison of the estimated energy of the gamma-candidate events extracted from the MAGIC standard analysis. Left panel shows the event-wise comparison, and the white-dashed line corresponds to $E_\mathrm{magic} = E_\mathrm{lst}$. Right panel shows the distribution of the relative differences of the estimated energies. The orange line corresponds to the result of the Gaussian fit around the peak of the distribution.}
  \label{energy_comparison}
\end{figure}
After the independent analyses, the coincident events are searched by using the developed algorithm and the following MAGIC standard analysis cuts are applied to extract gamma-candidate events. The first one is the so-called Hadronness cut ($h$ < 0.2), which is the output of a RF classifier with values ranging from 0 (very gamma-like events) to 1 (very hadron-like events). The other is the so-called Theta2 cut ($\theta^2$ < 0.025 deg$^2$), which is the squared angular distance between true and reconstructed arrival direction of the events. After the cuts, in total 1360 gamma-candidates are extracted from the coincident events. The expected contamination in the selected samples from background cosmic-ray events is $\sim$20\%. Finally, the LST-1 estimated energy of the gamma-candidates is compared with MAGIC one, and the results are shown in Fig.~\ref{energy_comparison}. The event-wise comparison in the left panel shows that the estimated energy of both systems is well correlated. Furthermore, the mean discrepancy is estimated to be 5\% $\pm$ 1\% of the MAGIC estimated energy as obtained by fitting a Gaussian around the peak of the relative difference distribution (right panel in Fig.~\ref{energy_comparison}). Considering the systematic uncertainty of the MAGIC energy scale \cite{MAGICper}, this result indicates that the accuracy of the LST-1 energy estimation is comparable to that of the MAGIC estimation. Also, the standard deviation of the fitted Gaussian $\sim$30\% is consistent with an expectation based on the fact that the energy resolution of each telescope system is $\sim$20\% (i.e., 20\% $\times \sqrt{2} \sim $30\%).

\section{Combined analysis of Crab Nebula data}
We have developed a pipeline dedicated to perform the combined analysis of LST-1 and MAGIC events. It reconstructs the geometrical shower parameters using the three shower images observed by LST-1, MAGIC-I and MAGIC-II, which improves the accuracy of the reconstruction compared to that using only the two MAGIC telescopes. Among those parameters, the $H_\mathrm{max}$ parameter, which represents the height of the shower maximum, is powerful for separating the gamma-hadron separation. Also, the Impact parameter, which represents the closest distance between the telescope and the reconstructed shower axis, is powerful for estimating the energy and arrival direction of the gamma rays \cite{MAGICstereo}. In addition, we have developed the MC simulation for the combined analysis of the joint observation data. We implemented the MAGIC telescopes into the \texttt{sim\_telarray} package, and the MAGIC telescope response is tuned to reproduce the standard one \cite{magic_lst_per}. 
\par
\begin{figure}
  \centering
  \includegraphics[width=15cm]{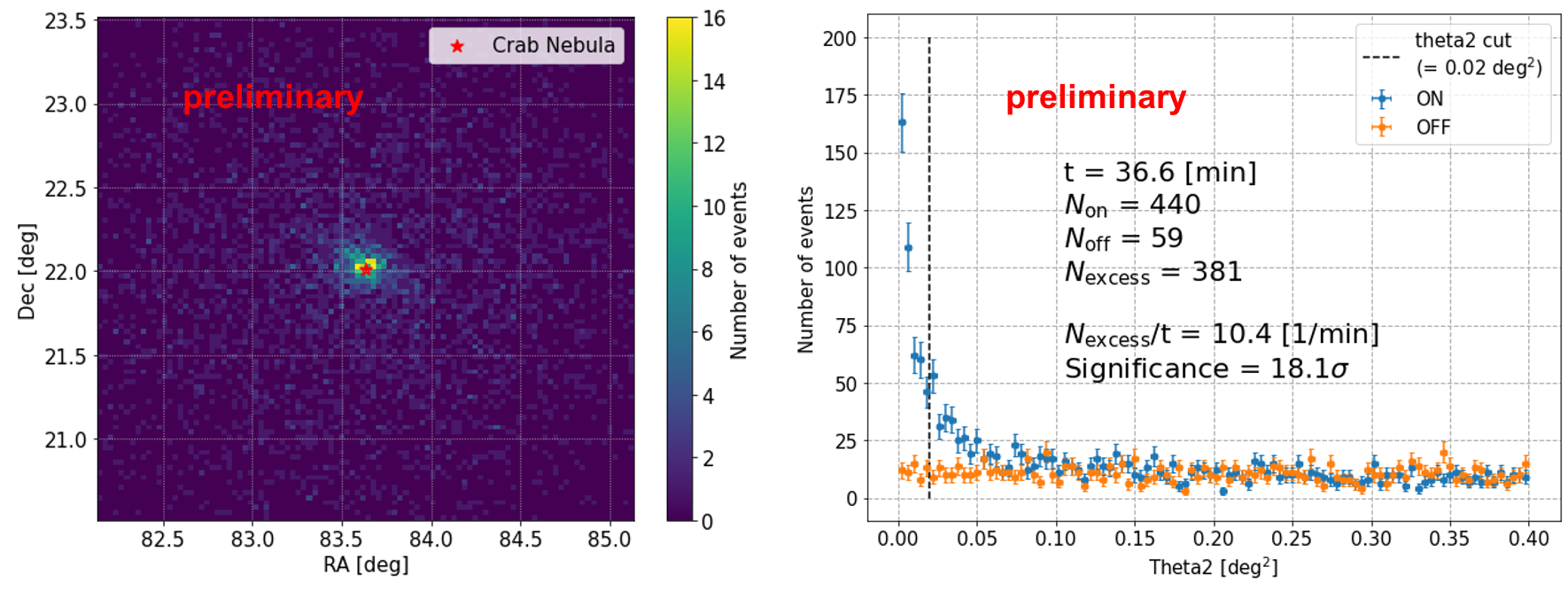}
  \caption{Signal from the Crab Nebula after the quality cuts, i.e., LST-1 events brighter than 100 photoelectrons and MAGIC events brighter than 50 photoelectrons. Left panel: Skymap of the Crab Nebula with the events surviving the Hadronness cut ($h$ < 0.2). Red star represents the position of the Crab Nebula. Right panel: Theta2 distribution with the events surviving the Hadronness cut ($h$ < 0.2). The blue plots represent the signal from the Crab Nebula position, and the orange plots represents the estimated background events. The black-dashed line represents the Theta2 cut which is applied to calculate the significance of the signal.}
  \label{crab_detection}
\end{figure}
\begin{figure}
  \centering
  \includegraphics[width=12cm]{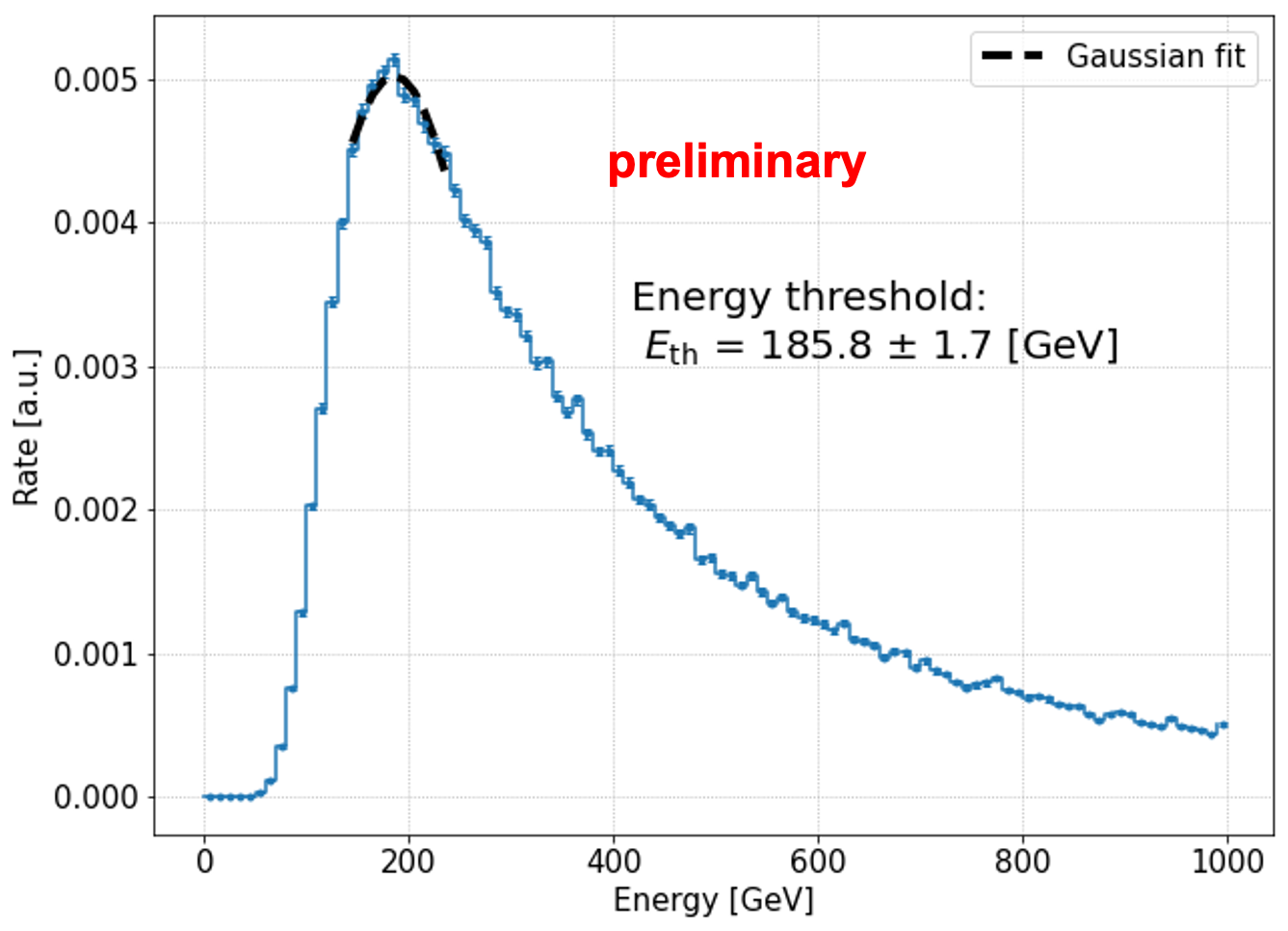}
  \caption{Rate of the MC gamma-ray events surviving the Hadronness cuts ($h$ < 0.2) and the quality cuts, i.e., LST-1 events brighter than 100 photoelectrons and MAGIC events brighter than 50 photoelectrons. The black-dashed line indicates the result of the Gaussian fit around the peak. The energy threshold $E_\mathrm{th}$ is defined as the mean value of the fitted Gaussian.}
  \label{energy_threshold}
\end{figure}
We present the preliminary results of the combined analysis of Crab Nebula data, taken on 18th November 2020 at middle zenith angles from 35 to 45 degrees. The data were taken with the wobble mode by both telescope systems with the same offset of 0.4 degree, which is suitable for performing the combined analysis. At first, the event coincidence algorithm is applied to the data to extract the events triggering the three telescopes, and the geometrical shower parameters are reconstructed with the three shower images. Then, the energy, arrival direction and Hadronness are reconstructed with the RF method, trained with the MC data produced with the similar zenith and azimuth angle as the real data. Finally, quality cuts are applied to the events based on the brightness of the images, i.e., we select the LST-1 events brighter than 100 photoelectrons and the MAGIC events brighter than 50 photoelectrons, different thresholds due to the areas of the reflective mirrors. The signal from the Crab Nebula after the Hadronness cut ($h$~<~0.2) is shown in Fig.~\ref{crab_detection}. The skymap in the left panel shows a clear excess from the direction of the Crab Nebula. Furthermore, the Theta2 distribution in the right panel shows that the Crab Nebula is detected with the statistical significance of 18.1~$\sigma$ with an observation time of 36.6 minutes. This is the first time that an astronomical source is detected with the combined analysis of LST-1 and MAGIC data. The number of background events is estimated from an OFF position defined on the opposite side of the Crab Nebula position with the same offset of 0.4 degree from the center of FoV. We estimated the energy threshold to be $\sim$185~GeV, defined as the energy where the trigger rate of MC gamma-ray events surviving the gamma-hadron separation cuts used in this analysis becomes maximum as shown in Fig.~\ref{energy_threshold}.

\section{Conclusion}
In this work we presented the analysis technique and the results of the inter-telescope cross-calibration and the combined analysis of LST-1 and MAGIC data. We have developed the event coincidence algorithm with timestamps in order to find the events triggering both telescope systems. The time-related parameters used in the algorithm are optimized, and it is confirmed that the coincidence rate of the cosmic-rays is consistent with that expected in the MC simulation. Then, we performed the cross-calibration based on the estimated energies of the coincident gamma-candidates, and it is shown that they are well correlated and the mean discrepancy is 5\% $\pm$ 1\%, which is of the same order of the systematic uncertainty of the MAGIC energy scale. We have also developed the pipeline for the combined analysis of the shower images observed by LST-1 and MAGIC. The preliminary results show that the Crab Nebula is detected above 185~GeV with the statistical significance of 18.1$\sigma$ with an observation time of 36.6 minutes, which is the first time that an astronomical source is detected with the combined analysis of LST-1 and MAGIC data. \par

\acknowledgments{We gratefully acknowledge financial support from the agencies and organizations listed here:\par
(CTA Consortium) http://www.cta-observatory.org/consortium\_acknowledgments\par
(MAGIC Collaboration) https://magic.mpp.mpg.de/acknowledgments\_ICRC2021}

\clearpage
\section*{Full Authors List:}

\subsection*{The CTA LST project:}
\scriptsize
\noindent
H.~Abe$^{1}$,
A.~Aguasca$^{2}$,
I.~Agudo$^{3}$,
L.~A.~Antonelli$^{4}$,
C.~Aramo$^{5}$,
T.~Armstrong$^{6}$,
M.~Artero$^{7}$,
K.~Asano$^{1}$,
H.~Ashkar$^{8}$,
P.~Aubert$^{9}$
% ...
\\

\noindent
$^{1}$Institute for Cosmic Ray Research, University of Tokyo.
$^{2}$Departament de Física Quàntica i Astrofísica, Institut de Ciències del Cosmos, Universitat de Barcelona, IEEC-UB.
$^{3}$Instituto de Astrofísica de Andalucía-CSIC.
$^{4}$INAF - Osservatorio Astronomico di Roma.
$^{5}$INFN Sezione di Napoli.
$^{6}$Aix Marseille Univ, CNRS/IN2P3, CPPM.
$^{7}$Institut de Fisica d'Altes Energies (IFAE), The Barcelona Institute of Science and Technology.
$^{8}$IRFU, CEA, Université Paris-Saclay.
$^{9}$LAPP, Univ. Grenoble Alpes, Univ. Savoie Mont Blanc, CNRS-IN2P3, Annecy.
% ...

\subsection*{The MAGIC Collaboration:}
\scriptsize
\noindent
V.~A.~Acciari$^{1}$,
S.~Ansoldi$^{2,41}$,
L.~A.~Antonelli$^{3}$,
A.~Arbet Engels$^{4}$,
M.~Artero$^{5}$,
K.~Asano$^{6}$,
D.~Baack$^{7}$,
A.~Babi\'c$^{8}$,
A.~Baquero$^{9}$,
U.~Barres de Almeida$^{10}$
% ...
\\

\noindent
$^{1}$ {Instituto de Astrof\'isica de Canarias and Dpto. de  Astrof\'isica, Universidad de La Laguna, E-38200, La Laguna, Tenerife, Spain} 
$^{2}$ {Universit\`a di Udine and INFN Trieste, I-33100 Udine, Italy} 
$^{3}$ {National Institute for Astrophysics (INAF), I-00136 Rome, Italy} 
$^{4}$ {ETH Z\"urich, CH-8093 Z\"urich, Switzerland} 
$^{5}$ {Institut de F\'isica d'Altes Energies (IFAE), The Barcelona Institute of Science and Technology (BIST), E-08193 Bellaterra (Barcelona), Spain} 
$^{6}$ {Japanese MAGIC Group: Institute for Cosmic Ray Research (ICRR), The University of Tokyo, Kashiwa, 277-8582 Chiba, Japan} 
$^{7}$ {Technische Universit\"at Dortmund, D-44221 Dortmund, Germany} 
$^{8}$ {Croatian MAGIC Group: University of Zagreb, Faculty of Electrical Engineering and Computing (FER), 10000 Zagreb, Croatia} 
$^{9}$ {IPARCOS Institute and EMFTEL Department, Universidad Complutense de Madrid, E-28040 Madrid, Spain} 
$^{10}$ {Centro Brasileiro de Pesquisas F\'isicas (CBPF), 22290-180 URCA, Rio de Janeiro (RJ), Brazil} 
% ...
$^{41}$ {also at International Center for Relativistic Astrophysics (ICRA), Rome, Italy} 
% ...

\end{document}